\documentclass{article}

\usepackage{arxiv}

\usepackage[utf8]{inputenc} 
\usepackage[T1]{fontenc}    
\usepackage{hyperref}       
\usepackage{url}            
\usepackage{booktabs}       
\usepackage{amsfonts}       
\usepackage{nicefrac}       
\usepackage{microtype}      
\usepackage{cleveref}       
\usepackage{lipsum}         
\usepackage{graphicx}
\usepackage[numbers]{natbib}
\usepackage{doi}

\usepackage{forest}
\usetikzlibrary{shadows.blur}
\usetikzlibrary{arrows.meta,shapes,positioning,shadows,trees}
\usepackage{rotating}
\usepackage{lscape}
\usepackage{tabularx}
\usepackage{longtable,pdflscape,booktabs}
\usepackage{caption}
\usepackage{array}

\title{Playful but Persuasive: Deceptive Designs and Advertising Strategies in Popular Mobile Apps for Children}

    \author{Hannah Krahl\\
    ORCID: 0009-0001-1852-7180\\
    ATHENE Center, Technical University of Darmstadt\\
    Rheinstraße 75, 64295 Darmstadt, Germany\\
    \texttt{hannah.krahl@stud.tu-darmstadt.de}\\
    \AND
    Katrin Hartwig \\
    ORCID: 0000-0003-4875-0110\\
	Science and Technology for Peace and Security (PEASEC)\\
	Technical University of Darmstadt\\
	Pankratiusstraße 2, 64285 Darmstadt, Germany \\
	\texttt{hartwig@peasec.tu-darmstadt.de} \\
	\AND
    Ann-Kathrin Fischer\\
    ORCID: 0009-0000-4142-2814\\
    Institute of Psychology, Technical University of Darmstadt\\
    Alexanderstraße 10, 64283 Darmstadt, Germany\\
    \texttt{ann-kathrin.fischer@stud.tu-darmstadt.de}\\
    \AND
    Theodora Nikolakopoulou\\
    ORCID: 0009-0007-1660-5564\\
    Institute of Psychology, Technical University of Darmstadt\\
    Alexanderstraße 10, 64283 Darmstadt, Germany\\
    \texttt{theodora.nikolakopoulou@stud.tu-darmstadt.de}\\
    \AND
    Guy Pires Cabritas\\
    ORCID: 0009-0007-0407-783X\\
    Institute of Psychology, Technical University of Darmstadt\\
    Alexanderstraße 10, 64283 Darmstadt, Germany\\
    \texttt{guy.pirescabrita@gmail.com}\\
    \AND
    Eva Ungeheuer\\
    ORCID: 0009-0005-9453-4496\\
    Institute of Psychology, Technical University of Darmstadt\\
    Alexanderstraße 10, 64283 Darmstadt, Germany\\
    \texttt{eva.ungeheuer@stud.tu-darmstadt.de}\\
    \AND
    Nina Gerber\\
    ORCID: 0000-0001-9669-7276\\
    ATHENE Center, Technical University of Darmstadt\\
    Rheinstraße 75, 64295 Darmstadt, Germany\\
    \texttt{n.gerber@psychologie.tu-darmstadt.de}\\
    \AND
    Alina Stöver\\
    ORCID: 0009-0006-0941-595X\\
    ATHENE Center, Technical University of Darmstadt\\
    Rheinstraße 75, 64295 Darmstadt, Germany\\
    \texttt{stoever@psychologie.tu-darmstadt.de}\\
}

\renewcommand{\shorttitle}{Playful but Persuasive}

\hypersetup{
pdftitle={A template for the arxiv style},
pdfsubject={q-bio.NC, q-bio.QM},
pdfauthor={David S.~Hippocampus, Elias D.~Striatum},
pdfkeywords={First keyword, Second keyword, More},
}

\begin{document}
\maketitle

\begin{abstract}
Mobile gaming apps are woven into children’s daily lives. Given their ongoing cognitive and emotional development, children are especially vulnerable and depend on designs that safeguard their well-being. When apps feature manipulative interfaces or heavy advertising, they may exert undue influence on young users, contributing to prolonged screen time, disrupted self-regulation, and accidental in-app purchases. 
In this study, we examined 20 popular, free-to-download children's apps in German-speaking regions to assess the prevalence of deceptive design patterns and advertising. Despite platform policies and EU frameworks like the General Data Protection Regulation and the Digital Services Act, every app contained interface manipulations intended to nudge, confuse, or pressure young users, averaging nearly six distinct deceptive patterns per app. Most also displayed high volumes of non-skippable ads, frequently embedded within core gameplay. These findings indicate a systemic failure of existing safeguards and call for stronger regulation, greater platform accountability, and child-centered design standards.

\end{abstract}


\begin{figure*}
  \includegraphics[width=\textwidth]{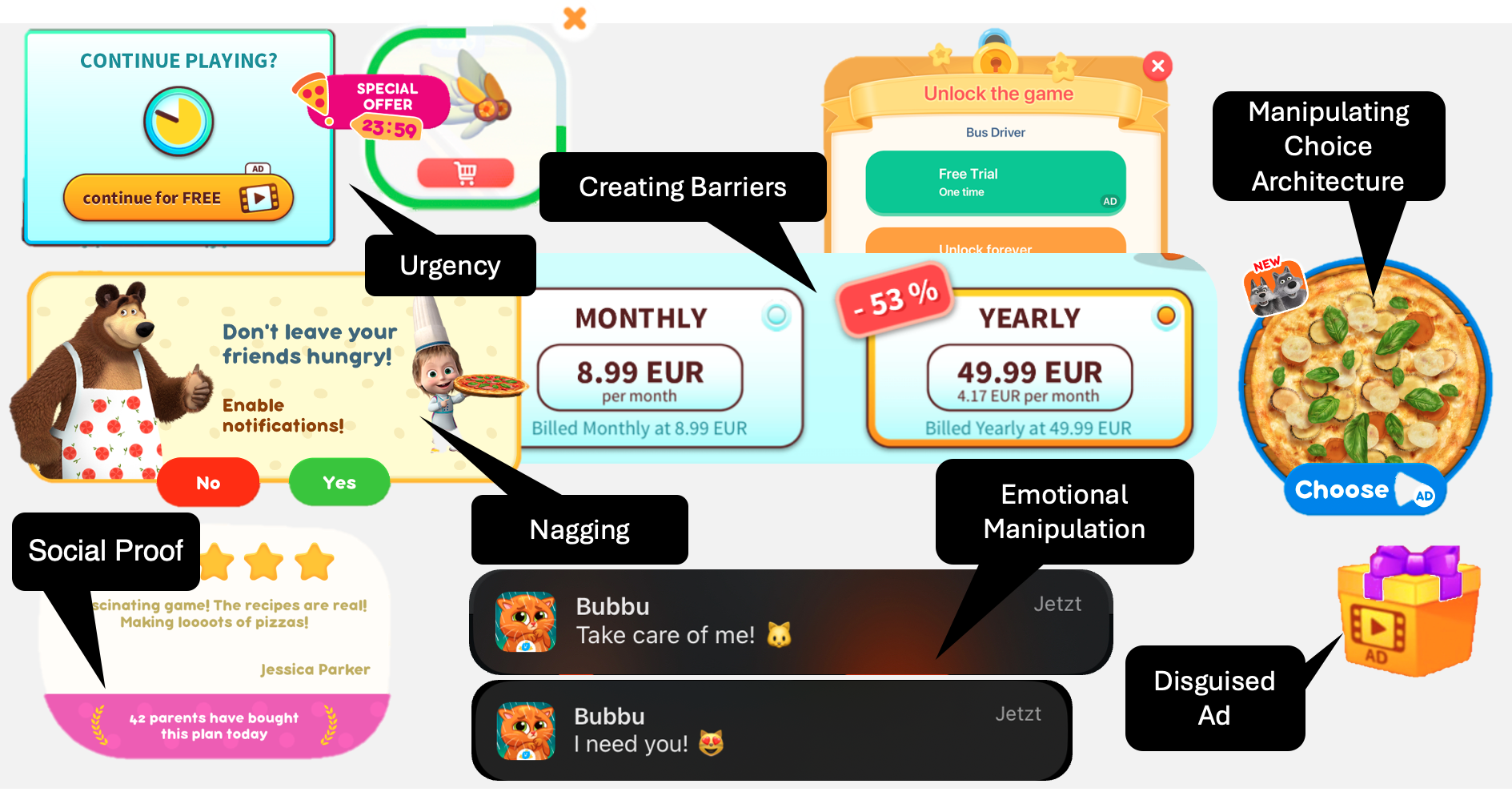}
  \caption{Compilation of screenshots of deceptive designs found in children's mobile gaming apps in our study.}
  \label{fig:overview}
\end{figure*}

\section{Introduction}
\label{Intro}
\textit{Imagine, after losing in a mobile game, a countdown appears, giving you mere seconds to revive and continue playing. The offer feels urgent and rewarding, yet the cost is hidden behind vague prompts: watch an ad or buy the premium package.} This is just one of many examples (see also Figure~\ref{fig:overview}) of deceptive designs commonly encountered in mobile games. Such scenarios capture the core nature of these patterns: interface elements deliberately crafted to steer users toward choices they might not have otherwise made \cite{Brignull2022}. Despite regulatory efforts, such as the EU Digital Services Act (DSA) or General Data Protection Regulation (GDPR), deceptive designs remain widespread across digital environments, including websites, social media, and mobile apps \cite{gunawan2021comparative, niknejad2024levelup, stover2022website}. For instance, they are frequently used in cookie banners, where users are nudged into giving consent through visual emphasis on acceptance options or by making rejection more difficult \cite{stover2022website, krisam2021dark}. In mobile games, these techniques are used to prolong playtime and increase monetization, often through slowed progression or time-based rewards that pressure players to spend either time or money \cite{Chamorro2024, radesky2022prevalence}.

While these practices raise concerns for all users, they are especially problematic when present in games designed for or accessible to children. Children are still developing the cognitive abilities needed to critically evaluate persuasive interfaces \cite{radesky2022prevalence}. They struggle to recognize commercial intent, resist impulses, and anticipate long-term consequences compared to adults \cite{Chamorro2024, renaud2024gullible}. Yet the youngest category in Apple’s App Store explicitly targets children aged 4+, and Google Play labels apps as “suitable for all ages”. This means that even preschool-aged children regularly encounter persuasive and commercialized environments in their everyday play.

From an HCI perspective, these concerns are central. Interfaces mediate user agency by shaping how choices are presented, constrained, and exercised, yet HCI’s understanding of autonomy and agency remains ambiguous, underscoring the need for conceptual clarity when studying manipulative designs \cite{bennett2023agency}. For children, interaction design intersects directly with ethics, as deceptive techniques exploit developmental vulnerabilities. At the same time, HCI research has shown how design can also empower children’s digital autonomy by scaffolding informed decision-making and supporting agency \cite{wang2023autonomy}. Systematically documenting these practices is therefore critical to advancing HCI scholarship and informing evidence-based game design standards and policy.

Monetized game mechanics, advertising, and affective or sensory manipulation are not deployed independently but are structurally intertwined within mobile game interaction flows \cite{sousa2023dark, radesky2022prevalence}. Advertising is frequently embedded as a forced or rewarded action at critical gameplay moments, while aesthetic and emotional cues amplify engagement and increase the effectiveness of monetization strategies. This interdependence motivates the need for an integrated perspective on the entire persuasive ecosystem children face in mobile games. In addition, there is a lack of empirical data on German-speaking markets and on the youngest age group (4+), even though these apps have been downloaded tens of millions of times and regulatory measures such as the DSA should be coming into force. 

In this paper, we directly address this gap by investigating the prevalence and the interplay of deceptive designs and advertising strategies in mobile games targeted at the youngest allowed age group (4+). Specifically, we seek to answer the following research question:

\vspace{1em}
What types of (a) deceptive designs and (b) advertising strategies are present in popular, free-to-download mobile gaming apps targeted at children and (c) how do these elements co-occur?
\vspace{1em}

To this end, we systematically analyzed 20 popular, free-to-download apps recommended for young children in German-speaking regions. Every single app contained multiple deceptive design patterns, with a median of 5.5 patterns per app and nearly all (90\%) featured extensive advertising. Importantly, advertising did not appear as an isolated layer but was tightly interwoven with core interaction mechanics and sensory cues. Across the sample, ads systematically overlapped with deceptive design patterns such as \textit{Forced Action}, where progression is contingent on engaging with advertising, and \textit{Sneaking}, where commercial content is embedded directly into gameplay \cite{gray2024ontology}. By tracing this joint occurrence of advertising, game mechanics, and sensory stimulation, our analysis moves beyond isolated pattern counts and shows that existing safeguards are failing, and provides an empirical foundation for an integrated assessment of children's apps. 

\vspace{1em}
\textbf{Contributions.} This paper makes three contributions to HCI research on deceptive game design in children’s digital environments. \textit{First}, it provides a systematic empirical analysis of the prevalence and distribution of deceptive design patterns and advertising practices in 20 popular, free-to-download mobile apps for the youngest target group (children aged 4+) in German-speaking markets. \textit{Second}, it advances prior work by examining how deceptive interface patterns, advertising formats, and sensory stimulation are not merely co-present but structurally intertwined within gameplay, revealing a tightly coupled persuasive ecosystem rather than isolated mechanisms. \textit{Third}, by situating these empirical findings within ongoing debates on user agency, child vulnerability, and platform responsibility, the paper contributes an evidence-based foundation for future HCI research, design guidelines, and regulatory discussions on child-centered app design.
   
\section{Related Work}
\label{RW}

This section provides an overview of deceptive designs\footnote{We use this term as an alternative to the still widely used `dark pattern', following recommendations of the ACM Diversity, Equity, and Inclusion Council and other organizations.}, focusing on their conceptualization, typologies, prevalence, and impact, particularly in the context of children's digital environments and advertising.

\subsection{Conceptualizing Deceptive Designs}

\textit{Deceptive designs} are defined as ``user interface design choices that benefit an online service by coercing, steering, or deceiving users into making unintended and potentially harmful decisions'' \cite[p.~2]{mathur2019dark}. \citet[p.~1]{Gray2018} emphasize that these practices reflect a structural shift in digital design where user value is systematically subordinated to shareholder value. Foundational work catalogs these strategies and shows their effects: \citet[]{bosch2016tales} introduce ``privacy dark strategies/patterns'' tied to cognitive biases, while \citet[]{luguri2021shining} show that mild to aggressive deceptive patterns can more than double, and sometimes nearly quadruple, opt-in rates for unwanted services.

Recent work shows that manipulative design is increasingly entangled with algorithmic decision-making and personalization, which can magnify deceptive effects \cite{Schafer2023Visual, hagendorff2024deception}. Such practices have been identified across multiple digital contexts, from websites and mobile applications to voice interfaces and extended reality environments \cite{owens2022ExploringInterfaces, krauss2024what}.

While much of this work frames deceptive designs as a consumer protection issue, HCI highlights its role as a design ethics problem. Interfaces mediate user agency by shaping how choices are structured and how autonomy is exercised, a concern reflected in recent work that both calls for clearer accounts of autonomy and agency in digital systems \cite{bennett2023agency} and demonstrates how design can actively support children’s digital autonomy \cite{wang2023autonomy}. This positions deceptive design not only as a regulatory concern but as a central challenge for HCI.

\subsection{Typologies of Deceptive Designs}

Several taxonomies have been developed to categorize deceptive design strategies, yet many differ in scope and terminology \cite{bosch2016tales, mathur2019dark, luguri2021shining}. To address this fragmentation, \citet{gray2024ontology} harmonized ten existing regulatory and academic frameworks into a comprehensive three-level ontology with standardized definitions for 64 design types. This ontology consists of:

\begin{enumerate}
\item \textbf{High-Level Patterns}: Broad strategies across platforms and media, such as \textit{Sneaking}, \textit{Obstruction}, \textit{Interface Interference}, \textit{Forced Action}, and \textit{Social Engineering}.
\item \textbf{Meso-Level Patterns}: Specific techniques bridging High-Level strategies and concrete implementations, such as \textit{Creating Barriers} as a form of the High-Level pattern \textit{Obstruction}.
\item \textbf{Low-Level Patterns}: Specific interface mechanisms directly linked to Meso-Level and High-Level strategies, such as \textit{Price Comparison Prevention} as a form of the Meso-Level pattern \textit{Creating Barriers}.
\end{enumerate}

This structured approach provides conceptual clarity for research and forms the basis of our analysis of deceptive design patterns in children’s apps.

\subsection{Advertising Strategies and Deceptive Designs}

Advertising and deceptive design practices are deeply intertwined, particularly in mobile gaming contexts aimed at children. Radesky et al. \cite{radesky2022prevalence} identify several recurring advertising strategies in children's apps, including \textit{Roadblock Ads} that interrupt gameplay with extended or forced interactions, \textit{Strategically Timed Ads} that appear at critical moments, and \textit{Ads with Reinforcement} that offer in-game rewards for viewing. These tactics frequently overlap with deceptive design patterns such as \textit{Forced Action}, where users must interact with an ad to continue, or \textit{Sneaking}, where commercial content is disguised as part of gameplay \cite{gray2024ontology}. 

Although ontologies of deceptive designs provide valuable frameworks for categorizing manipulative patterns \cite{gray2024ontology}, they often remain interface-centric and underplay the commercial logics rooted in advertising. In practice, however, advertising does not simply accompany deceptive design; it amplifies it by embedding monetization pressures directly into interaction flows. This entanglement is particularly relevant in children’s games, where persuasive ads and manipulative design strategies co-construct an immersive environment of influence that challenges children’s ability to exercise agency. Highlighting this overlap therefore helps bridge the gap between existing taxonomies of deceptive designs and the specific ways advertising shapes digital experiences for young users.

\subsection{Prevalence and Consequences of Deceptive Designs}

Despite regulatory efforts, deceptive designs remain pervasive across digital platforms, including e-commerce \cite{mathur2019dark}, cookie banners \cite{stover2022website}, and mobile applications \cite{diGeronimo2020UIPerception}. For instance, deceptive designs were identified in 95\% of top mobile apps \cite{diGeronimo2020UIPerception}. In gaming, these designs are strategically deployed to maximize retention and monetization \cite{niknejad2024levelup, king2018predatory}, often by exploiting cognitive biases and emotional triggers. Beyond such traditional platforms, recent HCI work shows that manipulative practices are also emerging in advanced contexts: conversational agents increasingly use deceptive tactics that exploit trust in AI systems \cite{scheurer2023llmdeception, hagendorff2024deception}, while XR environments create new opportunities for manipulation through embodied interactions \cite{krauss2024what}.  

The consequences of these practices are substantial. Even when users recognize manipulative designs, they frequently succumb to them \cite{bongard2021definitely}, and many view their reliance on digital services as unavoidable, especially when alternatives are lacking \cite{maier2020dark, acquisti2022nudges}. Taken together, these findings highlight not only the ubiquity of deceptive design but also its power to undermine agency, foster dependency, and exacerbate vulnerability.

\subsection{Children as a Target Group}

Children are uniquely vulnerable to deceptive designs due to developmental limitations in recognizing commercial intent, weaker impulse control, and heightened emotional susceptibility \cite{radesky2022prevalence, renaud2024gullible, Chamorro2024}. Empirical work shows that manipulative features are widespread in children’s apps and disproportionately affect children from lower-income households \cite{radesky2022prevalence}. Because these families are more likely to rely on free, ad-supported apps, their children are especially exposed to advertising-driven monetization practices that intertwine with deceptive interface design. Beyond mobile games, susceptibility to commercial influence is documented across diverse digital contexts, underscoring the need for child-centric design principles and policy safeguards \cite{moti2024children, LivingstoneStoilova2021}.

Interfaces structure choices and condition how autonomy is exercised; recent work calls for clearer accounts of autonomy and agency in digital systems \cite{bennett2023agency} and demonstrates how design can actively support children’s digital autonomy \cite{wang2023autonomy}. Examining deceptive practices in the German-speaking region, with its stringent regulatory framework (GDPR, DSA), therefore provides empirical grounding for aligning design practices with evidence-based regulation and child protection.
\section{Method}
\label{Method}

\begin{figure}[h]
  \centering
  \includegraphics[width=\textwidth]{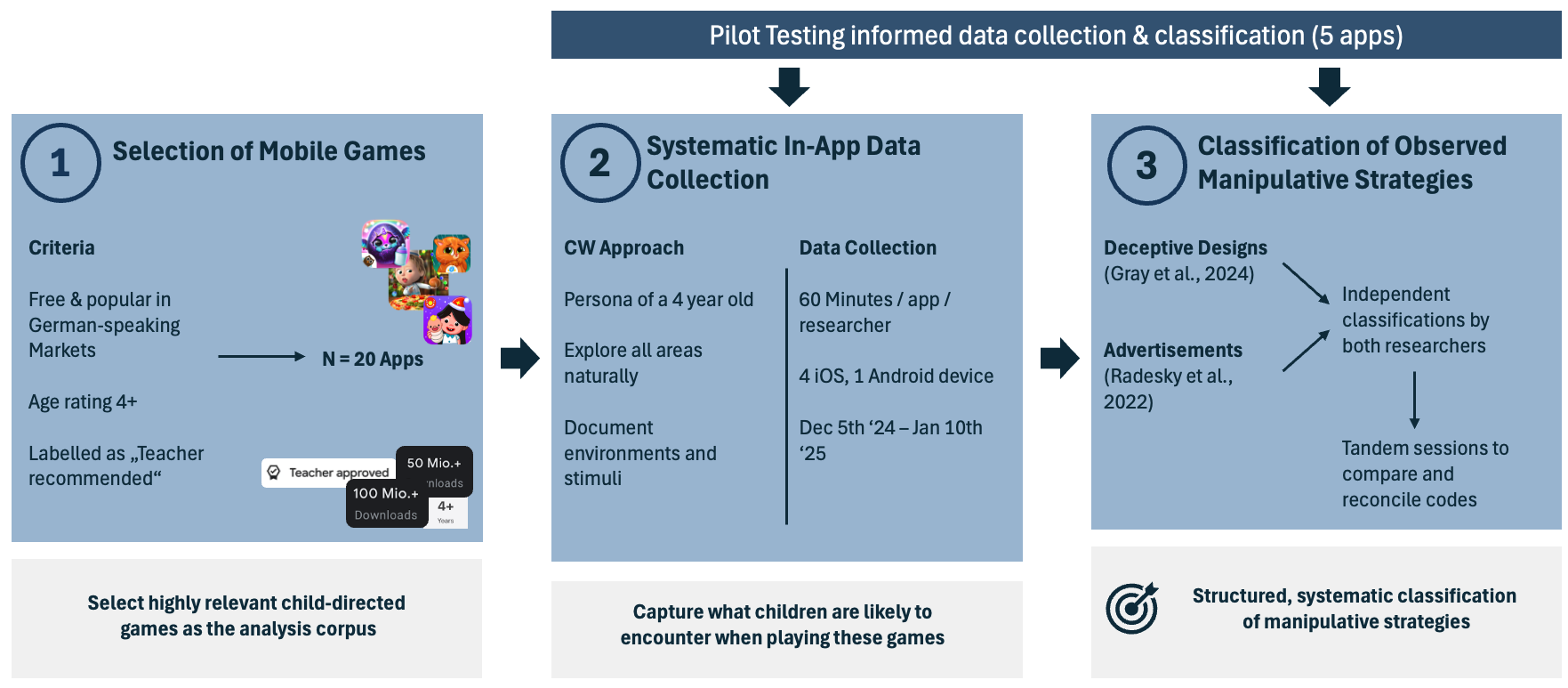} 
  \caption{Conceptual overview of the study design.}
  \label{fig:Method}
\end{figure}

This study investigated deceptive design patterns and advertising strategies in popular children's mobile apps. We followed a three-step process (Figure~\ref{fig:Method}).: (1) selecting relevant children's apps, (2) conducting a systematic in-app data collection via persona-based cognitive walkthroughs, and (3) classifying the observed deceptive patterns and advertising strategies using established typologies.

\subsection{Step 1: Selection of Popular Child-Directed Apps}
To identify the most popular children's mobile apps, we conducted a systematic screening using predefined inclusion and exclusion criteria.
Given the rapid turnover of mobile apps and the lack of transparent popularity metrics \cite[e.g., ][]{stufftv2025, applestore2025, sensortower2025}, we focused on free titles explicitly designed for children that were widely downloaded in the German-speaking market. To ensure popularity, relevance, and cross-platform consistency, we included only apps that: (i) were recommended for the youngest age group (rated 4+), (ii) had at least 50 million downloads, (iii) were available on both Android and iOS with identical titles and functionality, and (iv) were free to download. An initial screening using a 100-million-download threshold yielded only nine eligible apps; therefore, the threshold was lowered to 50 million downloads to obtain a sufficiently large yet analytically feasible sample. Because the iOS App Store does not publish download statistics, all popularity thresholds were based on Google Play metrics. Only apps available in identical versions on both platforms were included. The search and selection process took place on November 25, 2024.

The sample was drawn from the Google Play Store’s \textit{Kids} section, from which we selected only apps carrying the \textit{``Teacher Approved''} badge. This label indicates that, in addition to complying with Google's Family Policy, apps have undergone an additional review by a panel of teachers and child-development experts, assessing criteria such as age appropriateness, design quality, and suitability of advertising practices \cite{googleplay2024}. While this review focuses on child-appropriate content and advertising compliance, it does not explicitly address deceptive design practices.
The final sample comprised 20 apps listed in Tables~\ref{tab:patterns1} and~\ref{tab:patterns2}.

\subsection{Step 2: Systematic In-App Data Collection via Cognitive Walkthroughs}
In step 2, we collected data from the pre-selected 20 gaming apps by taking screenshots and manually taking notes following a systematic procedure. This allowed for the subsequent analysis and classification of deceptive patterns and advertising strategies in step 3. The data collection phase was embedded in persona-based cognitive walkthroughs to facilitate realistic encounters with the apps' interactions.

We employed a persona-based cognitive walkthrough (CW), while taking screenshots and taking notes in accordance with established principles of inspection-based evaluation \cite{wharton1994cognitive, mahatody2010cw}. The goal during data collection was not to assess usability or learnability, but to enable the systematic identification of how deceptive design patterns and advertising strategies become perceptible during interaction sequences, as many of these mechanisms unfold temporally and are contingent upon user actions.
The walkthrough adopted the perspective of a preschool child persona (approx. 4 years old), consistent with the 4+ age rating of all analyzed apps. This choice was theoretically grounded in prior work on early childhood touchscreen interaction, showing that young children exhibit lower touch precision and increased impulsive, exploratory tapping compared to adults \cite{vatavu2015touch, russojohnson2017alltapped}. These characteristics are directly relevant to how commercial prompts, forced actions, and misleading cues are encountered and acted upon.
In line with CW methodology, evaluators progressed through concrete interaction sequences and decision points, documenting at each step the perceived goal, available actions, and system feedback based on screenshots and additional manual notes. Because mobile games are inherently open-ended and non-linear, walkthroughs were structured around recurring in-game situations typical for children’s play (e.g., starting a level, receiving rewards, encountering locked content, responding to pop-ups), rather than fixed task scripts.

To approximate preschool behavior, evaluators followed a standardized protocol: (a) tap visually salient elements first, (b) repeat taps when feedback is delayed, (c) accept prompts and offers without reflective deliberation, (d) navigate opportunistically rather than strategically, and (e) treat surfaced ads as if they were part of the game.
Advanced gestures and parental control features were intentionally omitted, reflecting typical capabilities and access patterns in this age group. 

Each app was independently explored by two researchers for 60 minutes each during December 2024 and January 2025. During gameplay for data collection, all environments and relevant interface elements were systematically documented using screenshots and manual notes on the game's flow (e.g., when researchers encountered anomalies in the system's feedback, such as buttons not triggering a reaction). Researchers aimed to explore each app as comprehensively as possible by navigating through menus, participating in mini games, and interacting with in-game systems. Whenever new environments, design elements, or advertising content appeared, additional screenshots were captured. 

Before step 2 was conducted systematically, a \textit{pilot study} took place to assess the feasibility of our data collection and classification approach. Five apps meeting the selection criteria (see step 1) were independently analyzed by two researchers using CWs and the four-eyes principle \cite{kuepeli2024}. Playing the apps in parallel allowed for independent yet comparable observations and a shared understanding of how to classify deceptive design patterns. Pilot testing also informed procedural choices. A 60-minute playtime per app was set as a balance between depth and feasibility, as most deceptive patterns appeared early with diminishing returns over time. These insights supported the decision to analyze 20 apps, each reviewed by two researchers. While the sample represents only a fraction of the children’s app market, it includes highly popular titles likely to be encountered through app store discovery.

\subsection{Step 3: Classification of Deceptive Designs and Advertisements}
After data collection in step 2, both researchers independently classified all recorded environments and design elements according to the Meso-Level of \citet{gray2024ontology}'s ontology of deceptive design based on the screenshots and additional manual notes.
Our pilot confirmed the applicability of \citet{gray2024ontology} for coding deceptive patterns, with High- and Meso-Level categories proving most practical. Low-Level categorization was excluded due to unreliable mapping, highlighting challenges for future research. The pilot also showed that Gray’s ontology did not sufficiently capture advertising strategies. We therefore integrated the typology of \citet{radesky2022prevalence} and added two overarching categories: \textit{skippable} (ads that can be immediately dismissed) and \textit{non-skippable} (ads that require at least five seconds of viewing). This combined framework was used in the main study.
Designs exhibiting multiple patterns were coded under the most prominent category. Recurring instances of the same pattern within a single app were coded only once, regardless of their frequency. This approach was chosen to capture the diversity of deceptive patterns rather than their frequency of occurrence, as many design elements persisted continuously throughout gameplay or reappeared in identical form, making repeated frequency counts analytically uninformative in this context. The respective researcher pairs then compared and discussed findings. All screenshots were jointly reviewed to harmonize classifications, for example, when different gameplay paths revealed different design elements. In rare cases of unresolved disagreement regarding the most prominent deceptive pattern in a given design, classifications were discussed within the full research team, with screenshots serving as the primary basis for discussion and final decisions. 

For quantitative analysis, each deceptive design pattern was coded as a binary variable (1 = present during the 60-minute session; 0 = not observed). Advertising data were processed as absolute counts: across 120 minutes of gameplay per app (two independent sessions of 60 minutes), all advertisements were counted and classified according to \citet{radesky2022prevalence}, with the additional distinction between \textit{skippable} and \textit{non-skippable} ads. 

\subsection{Ethical Considerations}
The study aims to inform the regulation of harmful digital practices by documenting deceptive design patterns and advertising strategies in children’s apps. To limit the risk of misuse, only summarized results are shared publicly. More detailed findings are available upon justified request, ensuring transparency for researchers and regulators while discouraging opportunistic use by developers.  

No children took part in the study. All interactions were conducted by adult researchers simulating child-like gameplay behaviors. This allowed us to observe manipulative content without exposing minors to potential harm. Relying on trained adults also avoided the ethical and legal complexities of involving children, while still capturing realistic insights into how these mechanics operate in practice.
\section{Results}
\label{Results}

This section examines how frequently deceptive designs and advertising strategies occur in children’s apps and the various forms they take. Following \citet{gray2024ontology}, we use the distinction between High-Level and Meso-Level categories as a starting point for structuring our analysis, which then subsequently broadens into a more holistic evaluation.

\subsection{Prevalence and Classification of Deceptive Designs}

Deceptive designs were present in every app we analyzed (N=20). Across the dataset, we identified 116 instances, with a median of 5.5 distinct patterns per app (Table~\ref{tab:patterns2}). The distribution was highly uneven. Some popular games, such as \textit{Toca Boca World} and \textit{Avatar World}, incorporated only a few manipulative elements, while others relied heavily on them. \textit{Masha and the Bear: Pizzeria}, for example, stood out with 14 distinct patterns, making it the app with the highest density of deceptive designs.

\begin{figure*}[h]
  \centering
  \includegraphics[width=\textwidth]{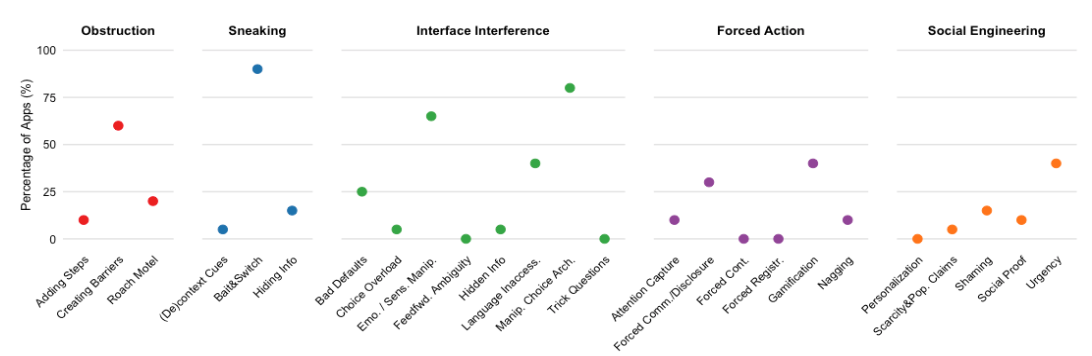} 
  \caption{Proportion of Apps in \% ($N = 20$) containing deceptive designs, categorized by High-Level (top labels, color sorted) and Meso-Level subtypes (bottom labels)}
  \label{fig:resultsalls}
\end{figure*}

The most universally present type was \textit{Interface Interference}: every app contained at least one example.
This often took the form of \textit{Manipulating Choice Architecture} (Figure~\ref{fig:resultsalls}), in which the interface was structured to steer children toward specific actions. In \textit{Masha and the Bear: Pizzeria}, for instance, attractive pizza options were locked behind advertising, while non-commercial alternatives were hidden from view (Figure~\ref{fig:overview}). Another frequent strategy was \textit{Emotional or Sensory Manipulation}(Figure~\ref{fig:resultsalls}), observed in 65\% of apps. Here, design elements used affective cues to pressure children into further engagement, as seen in \textit{Bubbu's Pet Cat}, where sad pet animations and emotionally loaded notifications urged users to return (Figure~\ref{fig:overview}).

\begin{table}[htbp]
\centering
\caption{Presence of deceptive design patterns in children's apps 1–10, based on the applied classification ontology by \citet{gray2024ontology}. A checkmark indicates that the respective pattern was identified in the app at least one time.}
\small
\begin{tabular}{p{1.5cm}p{4.3cm}*{10}{>{\centering\arraybackslash}p{0.5cm}}}
\toprule
\textbf{Category} & \textbf{Pattern} &
\rotatebox{90}{\textbf{Fluvsies}} & \rotatebox{90}{\textbf{Toca Boca World}} & \rotatebox{90}{\textbf{Pepi Hospital}} & \rotatebox{90}{\textbf{Bubbu Cat}} & 
\rotatebox{90}{\textbf{Barbie Dreamhouse}} & \rotatebox{90}{\textbf{Avatar World}} & \rotatebox{90}{\textbf{Panda Stadt}} & \rotatebox{90}{\textbf{Hair Salon 4}} & 
\rotatebox{90}{\textbf{Panda Schulbus}} & \rotatebox{90}{\textbf{Masha Games}} \\
\midrule
3*Obstruction
& Roach Motel & & & & \checkmark & & & & & & \\
& Creating Barriers & \checkmark & & \checkmark & \checkmark & \checkmark & & & & \checkmark & \checkmark \\
& Adding Steps & & & & \checkmark & & & & & & \\
 \midrule
3*Sneaking
& Bait and Switch & \checkmark & \checkmark & \checkmark & \checkmark & \checkmark & & \checkmark & \checkmark & \checkmark & \checkmark \\
& Hiding Information & & & \checkmark & & & & & & & \\
& (De)contextualizing Cues & & & & & & & & & & \\
 \midrule
8*Interface 
& Manipulating Choice Architecture & \checkmark & & \checkmark & \checkmark & \checkmark & \checkmark & & \checkmark & \checkmark & \\
Interference & Bad Defaults & & & \checkmark & & & & & \checkmark & & \checkmark \\
& Emotional or Sensory Manipulation & \checkmark & \checkmark & \checkmark & \checkmark & \checkmark & & & & \checkmark & \checkmark \\
& Trick Questions & & & & & & & & & & \\
& Choice Overload & & & & & & \checkmark & & & & \\
& Hidden Information & & & & & & & & & & \\
& Language Inaccessibility & & \checkmark & & & & \checkmark & \checkmark & & & \\
& Feedforward Ambiguity & & & & & & & & & & \\
 \midrule
6*Forced 
& Nagging & & & & & \checkmark & & \checkmark & & & \\
Interaction & Forced Continuity & & & & & & & & & & \\
& Forced Registration & & & & & & & & & & \\
& Forced Communication or Disclosure & & & & & & & \checkmark & & \checkmark & \\
& Gamification & \checkmark & & & & \checkmark & & \checkmark & & & \\
& Attention Capture & & & & & & & & & \checkmark & \\
 \midrule
5*Social 
& Scarcity and Popularity Claims & & & & & & & & & & \checkmark \\
Engineering & Social Proof & & & & & & & & & & \checkmark \\
& Urgency & \checkmark & & \checkmark & & & & & & \checkmark & \checkmark \\
& Shaming & & & & & & & & & \checkmark & \checkmark \\
& Personalization & & & & & & & & & & \\
\bottomrule
\end{tabular}
\label{tab:patterns1}
\end{table}

\begin{table}[htbp]
\centering
\caption{Presence of deceptive design patterns in children's apps 11–20, based on the applied classification ontology by \citet{gray2024ontology}. A checkmark indicates that the respective pattern was identified in the app at least one time.}
\small
\begin{tabular}{p{1.5cm}p{4.3cm}*{10}{>{\centering\arraybackslash}p{0.5cm}}}
\toprule
\textbf{Category} & \textbf{Pattern} &
\rotatebox{90}{\textbf{Masha Pizzeria}} & \rotatebox{90}{\textbf{Emily Erdbeer}} & \rotatebox{90}{\textbf{Hospital Stories}} & \rotatebox{90}{\textbf{Spielinsel}} & 
\rotatebox{90}{\textbf{My Town}} & \rotatebox{90}{\textbf{Panda Welt}} & \rotatebox{90}{\textbf{Bubbu School}} & \rotatebox{90}{\textbf{Paw Patrol}} & 
\rotatebox{90}{\textbf{Miga Stadt}} & \rotatebox{90}{\textbf{Pepi House}} \\
\midrule
3*Obstruction
& Roach Motel & & & \checkmark & & & & \checkmark & \checkmark & & \\
& Creating Barriers & \checkmark & \checkmark & & \checkmark & \checkmark & \checkmark & & \checkmark & & \\
& Adding Steps & \checkmark & & & & & & & & & \\
 \midrule
3*Sneaking
& Bait and Switch & \checkmark & & \checkmark & \checkmark & \checkmark & \checkmark & \checkmark & \checkmark & \checkmark & \checkmark \\
& Hiding Information & \checkmark & & & & & & & & \checkmark & \\
& (De)contextualizing Cues & & & & & & \checkmark & & & & \\
 \midrule
8*Interface 
& Manipulating Choice Architecture & \checkmark & \checkmark & \checkmark & \checkmark & \checkmark & \checkmark & & \checkmark & \checkmark & \checkmark \\
Interference & Bad Defaults & \checkmark & \checkmark & & & & \checkmark & & & & \\
& Emotional or Sensory Manipulation & \checkmark & & \checkmark & \checkmark & \checkmark & \checkmark & & & & \checkmark \\
& Trick Questions & & & & & & & & & & \\
& Choice Overload & & & & & & & & & & \\
& Hidden Information & & & & \checkmark & & & & & & \checkmark \\
& Language Inaccessibility & & & \checkmark & & \checkmark & \checkmark & \checkmark & \checkmark & & \\
& Feedforward Ambiguity & & & & & & & & & & \\
 \midrule
6*Forced 
& Nagging & & & & & & & & & & \\
Action & Forced Continuity & & & & & & & & & & \\
& Forced Registration & & & & & & & & & & \\
& Forced Communication or Disclosure & & \checkmark & \checkmark & \checkmark & & & & & \checkmark & \\
& Gamification & \checkmark & & \checkmark & \checkmark & \checkmark & & & & & \checkmark \\
& Attention Capture & \checkmark & & & & & & & & & \\
 \midrule
5*Social 
& Scarcity and Popularity Claims & \checkmark & & & & & & & & & \\
Engineering & Social Proof & \checkmark & & & & \checkmark & & & & & \\
& Urgency & \checkmark & & \checkmark & & \checkmark & & & & & \checkmark \\
& Shaming & \checkmark & & & & & & & & & \\
& Personalization & & & & & & & & & & \\
\bottomrule
\end{tabular}
\label{tab:patterns2}
\end{table}

\begin{figure}[h]
  \centering
  \includegraphics[width=\textwidth]{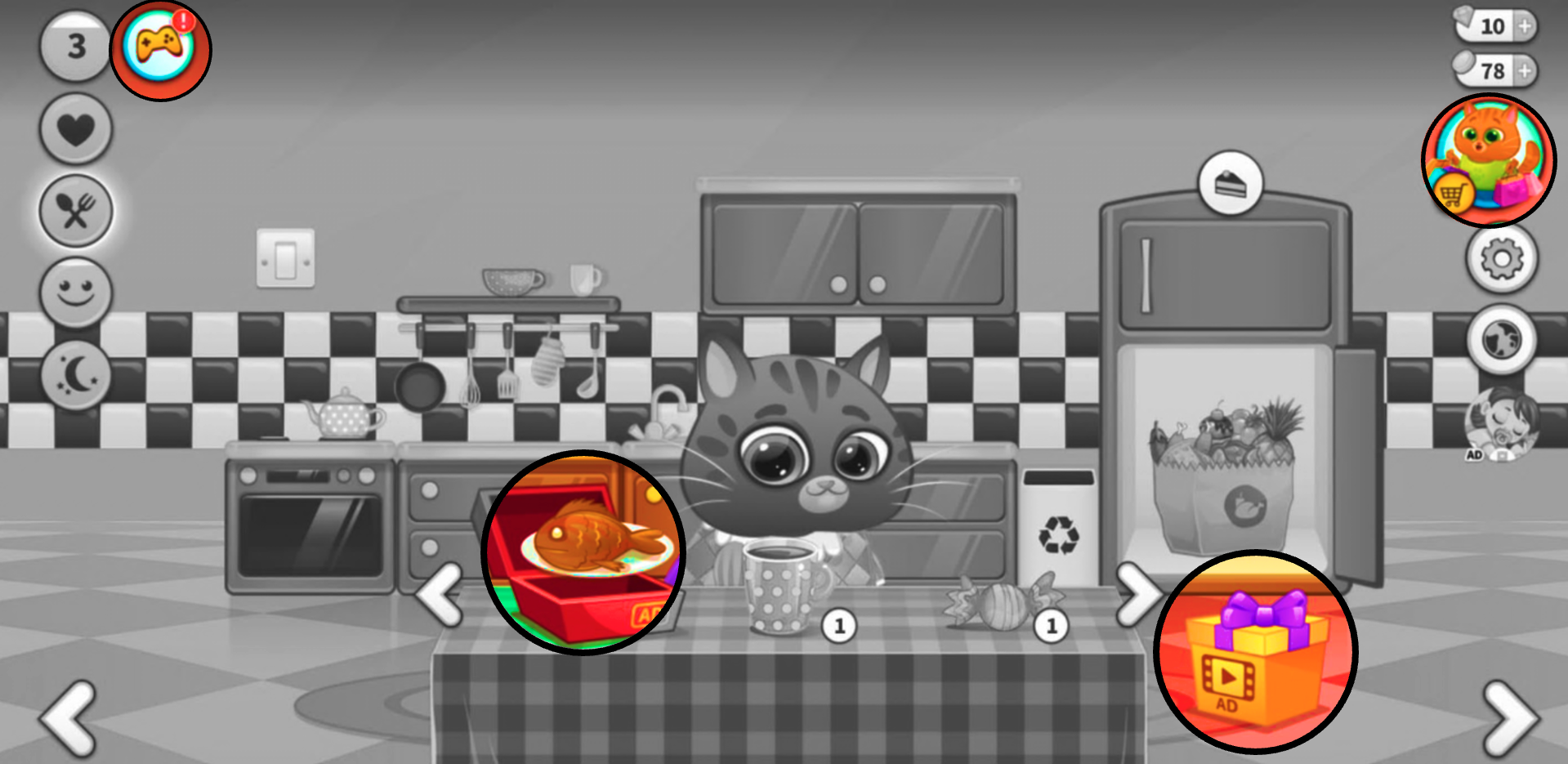} 
  \caption{Exemplary screenshot of deceptive designs in a children's app. 
(a) Top left: A jumping game button with sound effects entices players to click, triggering ads and \textit{Grinding} mechanisms (\textit{Forced Action}, \textit{Nagging}). 
(b) Middle left: The pet character demands to be fed with ad-locked items (\textit{Manipulating Choice Architecture}, \textit{Bad Defaults}). 
(c) Bottom right: A gift is offered in exchange for watching ads (\textit{Sneaking/Disguised Ad}, \textit{Visual Prominence}). 
(d) Top right: A pet store icon blows kisses to lure users into purchasing premium packages (\textit{Emotional/Sensory Manipulation}, \textit{Cuteness}).}
  \label{fig:screenshot}
\end{figure}

\textit{Sneaking} was identified in 90\% of apps, most often through the \textit{Bait and Switch} subtype, where ads were disguised as rewards (Figure~\ref{fig:resultsalls}). A typical example was a gift icon that, when tapped, launched an advertisement rather than providing in-game items (Figure~\ref{fig:screenshot}).

\textit{Obstruction} appeared in 70\% of apps, frequently in the form of \textit{Creating Barriers} (60\%), which complicated straightforward comparison of subscription prices. \textit{Masha and the Bear Games}, for instance, presented subscription tiers in ways that obscured clear cost evaluation (Figure~\ref{fig:overview}).

\textit{Forced Action} occurred in 55\% of the sample , typically through \textit{Gamification} mechanisms (40\%, (Figure~\ref{fig:resultsalls})). Many games required repetitive “grinding” in order to progress without payment, creating strong incentives to purchase shortcuts (Figure \ref{fig:screenshot}).

Finally, \textit{Social Engineering} was found in 40\% of apps, often through \textit{Urgency} tactics, such as limited-time offers that created artificial time pressure to trigger purchases (Figure~\ref{fig:overview}).

\subsection{Prevalence of Advertising Strategies}

\subsubsection{Frequency and Types of Ads}
Advertisements were nearly ubiquitous, with 90\% of apps displaying them. Across 40 hours of observation, we documented 500 advertisements, which we categorized following \citet{radesky2022prevalence} into three main types: \textit{Roadblock Ads}, which interrupt gameplay and prevent progress until the ad finishes; \textit{Strategically Timed Ads}, which appear at key moments such as the end of a level; and \textit{Ads with Reinforcement}, where users are rewarded with in-game benefits for watching. The distribution was as follows:
\begin{itemize}
\item 225 \textit{Roadblock Ads} (median = 7 per app)  
\item 85 \textit{Strategically Timed Ads} (median = 2.5 per app)  
\item 190 \textit{Ads with Reinforcement} (median = 2 per app)  
\end{itemize}
The majority of ads (72.8\%) were non-skippable, requiring complete viewing before gameplay could resume.

\subsubsection{Integration and Impact of Advertisements}
The intensity of advertising varied substantially. Eight apps contained more than 30 ads, another eight included fewer than ten, and only two apps were completely ad-free. \textit{Bubbu Pet Cat} displayed the highest number with 108 ads. Beyond sheer volume, integration practices made advertisements difficult to avoid or even recognize. Ads were embedded as banners, disguised through poorly labeled triggers, or presented as optional rewards, thereby blurring the boundary between play and commercial content. These practices not only disrupted gameplay but also forced repeated viewing in order to advance, significantly shaping the overall user experience and artificially extending session times.

\subsubsection{Child Safety Concerns with Advertising Content}
Beyond their frequency, advertisements raised clear safety concerns. Ads were often personalized based on presumed user preferences without consent, and in one instance exposed children to highly inappropriate material. An especially alarming case was an ad for a so-called \textit{“nude scanner”} app containing explicit imagery. Its appearance in a child-oriented game highlights how intrusive advertising normalizes unsafe interactions and exposes gaps in moderation and regulation.

\subsubsection{Sensory Overstimulation and Emotional Manipulation}
Many apps relied on intense sensory stimulation to hold children’s attention. This included continuous sound effects, rapidly changing animations such as bouncing buttons or flashing characters, and brightly colored pop-ups designed to divert focus. Emotional cues amplified these effects: childlike voices, sad or joyful character expressions, and persistent notifications encouraged ongoing engagement. These strategies created environments that were difficult to disengage from and risked fostering emotional dependency on the app.

\subsubsection{Content Access and Progression Barriers}
A common practice was to restrict access to core content or meaningful progression unless children either purchased premium features or repeatedly watched advertisements. Grinding mechanisms were frequently the only alternative, requiring extensive repetitive play to make minimal progress. Commercial prompts were often delivered by the app’s main characters, who used emotionally charged appeals to promote purchases. For children who could not pay, this design effectively reduced play to a cycle of repeated advertisement viewing, reinforcing commercial exposure as the central condition for advancement.

\subsubsection{Gradual Introduction and Technical Issues}
Deceptive design patterns rarely appeared all at once. Instead, they were gradually introduced: early gameplay was relatively smooth and inviting, while advertising pressure, monetization prompts, and manipulative interface elements escalated as children invested more time. This gradual ramp-up made practices harder to recognize and increased the likelihood of sustained engagement. Technical shortcomings further undermined safeguards. Parental age verification processes could be bypassed through unlimited retries, and several privacy-related controls were non-functional, leaving children unprotected against escalating manipulative tactics.
\section{Discussion}

This study examined how deceptive design patterns and advertising strategies shape children’s experiences in mobile games. Our findings show that deceptive design is pervasive: every app contained multiple patterns, often dominated by High-Level categories such as \textit{Interface Interference} and \textit{Sneaking}. These practices appear to contradict platform policies, yet they remain widespread, indicating gaps in enforcement and monitoring.  

Advertising was tightly woven into gameplay and largely unavoidable. Across the sample, children were repeatedly exposed to non-skippable ads, strategically timed prompts, and even age-inappropriate content. This highlights potential shortcomings in current moderation and child-protection mechanisms, particularly given the prevalence of non-skippable and strategically timed advertising within the gameplay. The integration of manipulative advertising with deceptive design patterns, such as \textit{Obstruction} and \textit{Forced Action}, illustrates how monetization pressures create immersive environments of commercial influence.  

Together, these findings suggest that existing taxonomies should account for advertising practices more explicitly and highlight how current industry models exploit children’s cognitive and emotional vulnerabilities, rather than supporting transparency and ethics.

\subsection{Potential Harmful Consequences for Children}
The mobile games analyzed in our study heavily exploit children’s limited ability to recognize persuasive intent by embedding deceptive patterns directly into play. This leads to several potentially harmful consequences.

First, deceptive designs such as \textit{Emotional or Sensory Manipulation} and \textit{Bad Defaults} (\textit{Interface Interference}), \textit{Attention Capture} and \textit{Gamification} (\textit{Forced Action}), or \textit{Shaming} and \textit{Urgency} (\textit{Social Engineering}) artificially prolong children’s play time by creating pressure and incentives to remain engaged or return frequently. Pediatric guidelines recommend limiting preschoolers to no more than two hours of daily screen exposure \cite{AAP_media_2013}. Most children already exceed these limits \cite{tandon2011preschoolers}, and a recent meta-analysis found that only about one third of 2–5-year-olds meet the stricter recommendation of under one hour per day \cite{mcarthur2022global}. \citet{radesky2022prevalence} reported that about a quarter of children’s apps use characters to pressure users into continued play, undermining autonomy. Screen-time endings are often accompanied by negative emotions such as sadness, anger, and frustration \cite{read2025emotions}. Psychiatric research further shows significant associations between high smartphone use and depressive symptoms, conduct problems, and emotional difficulties in young people \cite{mulla2024screen}. 

Secondly, many apps in our sample relied on frequent notifications, sometimes even without explicit consent, or employed approval begs that themselves contained deceptive elements (e.g., ``Don’t leave your friends hungry! Enable notifications'', see Fig. \ref{fig:overview}). Such practices exacerbate the disruptive potential of notifications. Prior work has shown that even the sound or vibration of a notification reduces performance on attention-demanding tasks in adults \citet{wilmer2017smartphones}. Extending this, \citet{upshaw2022hidden} demonstrate that not only alerts but even the mere presence of a smartphone can impair attentional control, while \citet{castelo2025blocking} show that blocking mobile internet improves sustained attention. Given children’s more limited regulatory capacities, these effects are likely to be even more pronounced. Many of the apps we studied exploited this vulnerability by sending emotionally loaded push messages (e.g., ``I want to cuddle!'') or embedding parasocial characters to shame users into returning. Such tactics may interfere with attentional regulation and contribute to repeated engagement driven by affective pressures.

Third, monetization strategies were pervasive. We found that 60\% of apps used deceptive barriers such as price-comparison prevention or intermediate currencies, and 80\% manipulated choice architectures (Figure~\ref{fig:resultsalls}). \citet{aagaard2022game} report “numerous accounts of young children spending tens of thousands of US dollars on casual mobile games,” illustrating how monetization schemes target vulnerable users. High-profile cases underscore the systemic nature of these risks: Apple’s 32.5 million USD settlement for unauthorized in-app purchases \cite{appleFTC2014} and Epic Games’ 245 million USD FTC fine \cite{FTC2023Press} for deceptive interfaces. Tactics include obscuring costs through bundles or intermediate currencies, and using parasocial characters to nudge children into subscriptions or virtual gifts. Industry insiders label high-spending users as “whales”, openly acknowledging business models that exploit compulsive spending. Because children lack a mature understanding of persuasive intent, they are particularly susceptible. Supporting this, \citet{hardwick2025scamming} found that while children aged 7–14 often recognized deception, this very recognition heightened their sense of harm.

Fourth, beyond immediate attentional and financial harms, manipulative practices risk fostering emotional dependence, normalizing deceptive designs, and exposing users to excessive advertising. Children may experience discomfort or anxiety when unable to meet virtual demands, reinforcing cycles of compulsion. When even adults struggle to recognize deceptive practices \cite{bourdoucen2023}, such practices become especially dangerous for digitally inexperienced users. Repeated exposure to targeted advertising and manipulative interfaces can normalize these tactics, shaping users’ expectations of digital interaction. In our study, children were exposed to an average of 25 advertisements in just two hours of play, with 359 of the 500 being unskippable. Developmental research shows that children under six do not yet possess the critical reasoning skills needed to resist advertising, and even older children remain strongly influenced \cite{packer2022critical}. Exposure to such volumes of advertising may normalize commercial persuasion as part of play, hindering the development of digital literacy. Without interventions, children risk perceiving deceptive patterns as standard features of digital interaction, which may shape long-term expectations of digital interaction and challenge the development of meaningful consent.

\subsection{Societal Aspects and Digital Justice}

Free apps consistently include more manipulative design elements than paid ones. \citet{radesky2022prevalence} found that free apps contained a median of three manipulative features compared to one in paid apps, with children from lower-education households particularly exposed to purchase pressure. This dynamic may exacerbate digital inequality, as children from lower-income families are more likely to rely on free content. The pattern mirrors broader HCI findings on the distribution of deceptive designs: free-to-play games and apps are especially prone to exploitative monetization \cite{aagaard2022game}.

Parents often rely on app store labels such as ``teacher recommended'' \cite{googleplay2024}, or ``educational'' (Apple App Store), assuming these categories signal quality and safety. Yet content analyses show that many such apps are dominated by disruptive advertising and test-oriented feedback rather than meaningful learning. Prior CHI work notes that app ecosystems obscure manipulative practices, placing the burden on parents to detect them \cite{Gray2018}—a challenge given varying levels of digital literacy. Moreover, monetization strategies can shift over time: in our study, several apps began with minimal advertising but later adopted aggressive monetization, undermining one-off parental checks.

\subsection{Guidelines and Perspectives for Ethical Children's Game Design}
The findings of this study underscore the urgent need for ethical frameworks that can guide the design of children’s digital games and apps. While persuasive and monetization-driven design has become a dominant industry practice, the evidence presented here illustrates that such practices can undermine children’s autonomy, well-being, and trust. In line with recent HCI work that calls for accountability in deceptive design research and regulation \cite{gray2024ontology, fitton2019framework}, we propose several directions for ethical design.

\subsubsection{Transparency through labeling and rating systems}
\citet{aagaard2022game} suggested the introduction of a ``Dark Pattern Badge'', which would signal the presence and type of manipulative features in free-to-play games. We extend this proposal by recommending that such badges also quantify and categorize advertising practices, including frequency, intrusiveness, and skippability of ads. A transparent rating system could be integrated into app stores, giving parents accessible information about the persuasive strategies embedded in children’s apps. This would align with broader HCI efforts to develop evidence-based rating systems and platform accountability mechanisms.
 
\subsubsection{Educating developers on the impact of manipulative design decisions} 
Most deceptive features are introduced to maximize engagement or revenue, but many developers may not fully grasp the developmental and psychological consequences for child users. Research shows that manipulative design can prolong playtime, trigger negative emotions at transition points, encourage compulsive return behavior, and normalize commercial persuasion \cite{read2025emotions, radesky2022prevalence}. Integrating this knowledge into developer training, through workshops, certification programs, or HCI-informed design curricula, could help shift industry practices. Developers should be exposed to empirical evidence about how children respond differently from adults to persuasive cues and advertising, and why such vulnerabilities require special protection.

\subsubsection{Digital literacy for parents as protection against manipulative design} 
Parents should not rely solely on age ratings or app store categories but should actively engage with their children’s digital experiences. They should be encouraged and trained in digital literacy, which includes the ability to recognize deceptive design. Targeted awareness campaigns are needed to inform parents about manipulative practices and their ethical implications. Importantly, parents should remain cautious not only during the initial stages of gameplay but also through regular check-ins, as manipulative design features or higher volumes of advertising may be introduced over time.

\subsection{Limitations and Future Research}
This observational study provides valuable insights into deceptive design and advertising practices in children’s apps; however, several limitations should be noted. 

\textit{First}, the sample was limited to 20 free-to-play apps from the Google Play Store’s children’s section in German-speaking regions. While we chose a systematic approach of app selection, this excludes paid apps, iOS-exclusive titles, and region-restricted content, and may not reflect the global diversity of children’s apps. We want to emphasize that the reliance on Google’s ``Family-Friendly'' label may also have biased the sample toward content deemed appropriate, and popularity was assessed using Google Play metrics only. 

\textit{Second}, data were collected between December 2024 and January 2025. Given the pace of app updates and ongoing regulations (e.g., GDPR), the findings may not generalize across time or contexts and require further investigation. While our study provides a systematic starting point, future studies might observe the fast-paced changes over time.

\textit{Third}, gameplay was conducted by adults simulating child-like interaction styles. While this ensured systematic identification, it reduces ecological validity, as children may respond differently, and some manipulative elements could have gone undetected. Developing ethically sound participatory methods remains an important challenge for future research. 

\textit{Fourth}, each app was tested for 60 minutes. Manipulative features triggered only after extended or repeated play may therefore have been underestimated. Future studies may investigate deception during long-term app usage to address these potential additional effects.
\section{Conclusion}

This study provides the first integrated analysis of deceptive design and advertising strategies in popular children’s apps in German‑speaking markets. By examining how these elements intertwine, we show that existing safeguards are insufficient and that manipulative ecosystems are commonplace. While our sample is limited to 20 free‑to‑download apps and simulates child interaction through persona‑based walkthroughs, the consistent presence of deceptive practices highlights an urgent need for more stringent regulation and industry accountability. Future research should expand to larger and more diverse samples, include participatory studies with children, and investigate the long‑term effects of deceptive interfaces on child development.

\section{Acknowledgements}
This work was supported by the German Federal Ministry of Education and Research and the Hessian Ministry of Higher Education, Research, Science and the Arts within their joint support of the National Research Center for Applied Cybersecurity ATHENE.




\bibliographystyle{ACM-Reference-Format}
\bibliography{references}

\appendix
\section{Appendix}
\label{Appendix}
In this section, we provide further material and results from our study.
\begin{footnotesize}
\begin{longtable}{p{2.2cm}p{3.4cm}p{3.7cm}p{4.2cm}}
\caption{Ontology based on \citet{gray2024ontology}, including definitions of high-level and meso-level patterns that served as the basis for the classification of the patterns identified in our study.}
\label{tab:ontologygray}\\
\toprule
\textbf{High Level Pattern} & \textbf{Definition (High Level)} & \textbf{Meso Level Pattern} & \textbf{Definition (Meso Level)} \\
\midrule
\endfirsthead
\multicolumn{4}{c}{{\textit{Table \thetable\ (continued)}}}\\
\toprule
\textbf{High Level Pattern} & \textbf{Definition (High Level)} & \textbf{Meso Level Pattern} & \textbf{Definition (Meso Level)} \\
\midrule
\endhead
\multicolumn{4}{r}{{\textit{Continued on next page}}}\\
\endfoot
\endlastfoot
Sneaking & Sneaking hides, disguises, or delays the disclosure of important information that would otherwise influence the user to act differently. & Bait and Switch & Subverts the user’s expectation that their choice will result in a desired action, instead leading to an unexpected, undesirable outcome. \\
& & Hiding Information & Subverts the user’s expectation that all relevant information is available, instead hiding or delaying disclosure. \\
& & (De)contextualizing Cues & Confuses users or hides relevant info by placing it in misleading or inappropriate contexts. \\
\midrule
Obstruction & Obstruction impedes a user’s task flow, making actions more difficult than necessary to dissuade or deter users. & Roach Motel & Easy to enter, hard to exit—contrary to the expectation that actions are reversible. \\
& & Creating Barriers & Prevents or complicates a user task to discourage completion. \\
& & Adding Steps & Inserts unnecessary steps to make tasks harder, against expectations of simplicity. \\
\midrule
Interface Interference & Interface Interference manipulates the UI to privilege certain actions, confuse users, or hide better alternatives. & Manipulating Choice Architecture & Structures options to lead toward undesirable outcomes. \\
& & Bad Defaults & Uses pre-selected settings that are not in users’ best interest. \\
& & Emotional or Sensory Manipulation & Uses visual/auditory cues to steer emotions and behavior. \\
& & Trick Questions & Uses confusing language or layout to induce wrong choices. \\
& & Choice Overload & Overwhelms with too many options, hindering informed decision-making. \\
& & Hidden Information & Hides or disguises relevant information. \\
& & Language Inaccessibility & Uses language complexity to prevent informed decisions. \\
& & Feedforward Ambiguity & Mismatch between interface cues and outcomes. \\
\midrule
Forced Action & Forced Action requires users to perform unrelated or undesirable tasks to proceed, interrupting their experience. & Nagging & Repetitively interrupts the user to pressure a certain action. \\
& & Forced Continuity & Makes it hard to cancel or notice ongoing subscriptions. \\
& & Forced Registration & Coerces or tricks users into unnecessary sign-up. \\
& & Forced Communication or Disclosure & Induces oversharing or misuses shared information. \\
& & Gamification & Coerces users into continuous use misaligned with their goals. \\
& & Attention Capture & Traps users into staying longer than intended. \\
\midrule
Social Engineering & Social Engineering uses cognitive biases and social norms to manipulate behavior. & Scarcity or Popularity Claims & Pushes decisions with exaggerated or false urgency/popularity. \\
& & Social Proof & Suggests that others’ behavior is the correct choice. \\
& & Urgency & Emphasizes limited-time offers or deadlines, real or not. \\
& & Shaming & -- \\
& & Personalization & Uses personal data to steer decisions and hide alternatives. \\
\bottomrule
\end{longtable}
\end{footnotesize}



\end{document}